\documentclass[epj]{svjour}
\usepackage{graphicx}
\usepackage{amssymb}
\usepackage{amsmath}
\usepackage{times}

\newcommand{\ud}{\mathrm{d}}
\newcommand{\PRL}{Phys. Rev. Lett. }
\newcommand{\PRB}{Phys. Rev. B }
\newcommand{\APL}{Appl. Phys. Lett. }

\begin{document}
\title{Inelastic relaxation and noise temperature in S/N/S junctions}
\author{C. Hoffmann \and F. Lefloch \and M. Sanquer}                     % Do not remove

\institute{D\'epartement de Recherche Fondamentale sur la Mati\`ere Condens\'ee / SPSMS / LCP,  CEA-Grenoble,\\ 17 rue des Martyrs, 38054 Grenoble cedex 09, FRANCE}
\mail{hoffmann@drfmc.ceng.cea.fr}
%
%\date{Received: 5 February 2002 / Revised version: 28 June 2002}
\abstract{We studied electronic relaxation in long diffusive superconductor / normal metal / superconductor (S/N/S) junctions by means of current noise and transport measurements down to very low temperature ($100\, mK$). Samples with normal metal lengths of $4, 10$ and $60\, \mu m$ have been investigated. In all samples the shot noise increases very rapidly with the voltage. This is interpreted in terms of enhanced heating of the electron gas confined between the two S/N interfaces. Experimental results are analyzed quantitatively taking into account electron-phonon interaction and heat transfer through the S/N interfaces. Transport measurements reveal that in all samples the two S/N interfaces are connected incoherently, as shown by the reentrance of the resistance at low temperature. The complementarity of noise and transport measurements allows us to show that the energy dependence of the reentrance at low voltage is essentially due to the increasing effective temperature of the quasiparticles in the normal metal.  
\PACS{
      {74.50.+r}{Proximity effects, weak links, tunneling phenomena, and Josephson effects}   \and
      {74.80.Fp}{Point contacts; SN and SNS junctions}    \and
      {73.50.Td}{Noise processes and phenomena} 
     } % end of PACS codes
} %end of abstract
\maketitle
\begin{figure}[h]
\includegraphics[width=0.48\textwidth,bb=15 15 740 525,clip]{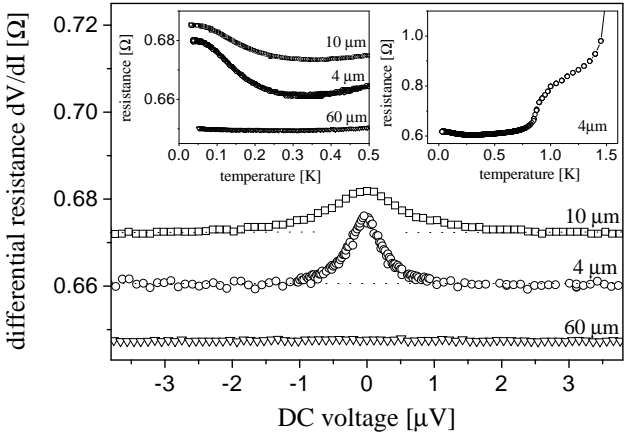}
\caption{Differential resistance $dV/dI$ versus DC voltage at $100\, mK$ for 3 samples with different lengths (data for the $60\, \mu m$ sample are shifted by $+0.06\, \Omega$). Right inset: Resistance versus temperature for a slightly different $4\, \mu m$ sample. Left inset: The resistance of the 3 samples at very low temperatures.}
\label{fig:r(v)}
\end{figure}
\begin{figure}[h]
\includegraphics[width=0.48\textwidth,bb=0 0 740 525,clip]{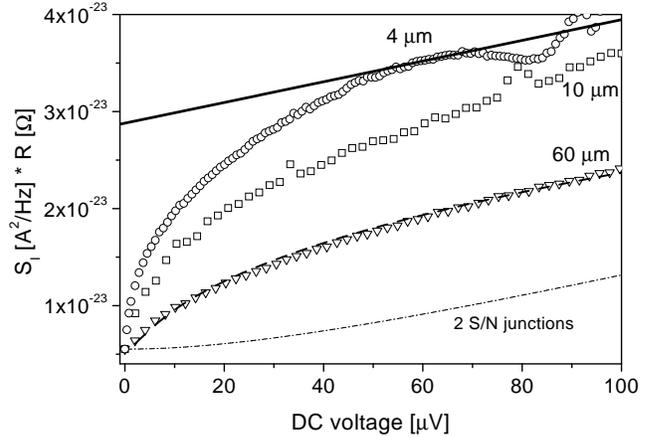}
\caption{Current noise density times the resistance $R=V/I$ versus DC voltage at $T=100\, mK$. Dot-dashed line: expected noise of 2 S/N junctions in series. Solid straight line: Theoretical prediction~\cite{Bezuglyi,Nagaev2001} in the collisionless regime with $\Delta =135\, \mu eV$. Dashed line: Fit taking into consideration e-ph interaction (see text).}
\label{fig:sr(v)}
\end{figure}
\begin{figure}[h]
\includegraphics[width=0.48\textwidth,bb=13 14 760 550,clip]{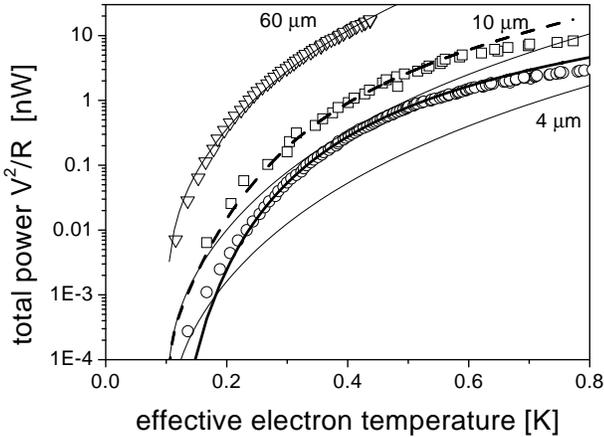}
\caption{Total power injected in the junctions as a function of the effective electron temperature in the normal part at a base temperature $T=100\, mK$ (symbols) and theoretical predictions: thin solid lines - power dissipated by e-ph-scattering according to Eq. (\ref{e-ph}) for the 3 samples respectively; thick solid line - power dissipated through the S/N interfaces according to Eq. (\ref{Pns}) for the $4\, \mu m$ sample; dashed line - sum of the contributions of the two cooling mechanisms for the $10\, \mu m$ sample.}
\label{fig:pns}
\end{figure}

The profound comprehension of the current transport in metals is a topic of permanent interest~\cite{LesHouches,Quirion}. With the progress in thin film technology a great number of studies deal with coherence phenomena at low temperature in metallic samples of length $L$ shorter than the phase coherence length $L_{\phi}$ (mesoscopic regime). In this context, a lot of works focus on inelastic processes and for instance, on the apparent saturation of the phase breaking length~\cite{Mohanty,Pierre}. Current noise measurement is particularly well suited to such investigations because it is sensitive to energy relaxation processes and gives access to the involved inelastic scattering length $L_{in}$~\cite{Naveh,Blanter}.\\
Current noise in diffusive mesoscopic normal metals connected to two normal reservoirs (N/N/N-case) has been studied by various groups~\cite{Steinbach,Schoelkopf,Henny}. Using a Boltzmann-Langevin approach the current noise is given by~\cite{Nagaev}:
\begin{equation}
S_I=\frac{4}{RL}\int_{-L/2}^{L/2} \ud x \int \ud \epsilon \, f(\epsilon ,x)[1-f(\epsilon ,x)], \label{SiNag1}
\end{equation}
where $f$ is the distribution function of the electrons and $R$ the resistance of the sample. In the regime $L<<L_{in}$, the noise is reduced by a factor 3 compared to the Schottky value $2eI$. If the length of the sample exceeds $L_{in}$, the electron gas can be described by a Fermi distribution with an effective temperature $T_e$ and Eq. (\ref{SiNag1}) simplifies to:
\begin{equation}
S_I=\frac{4k_B}{RL}\int_{-L/2}^{L/2} \ud x\,  T_e(x) = \frac{4k_B\overline{T_e}}{R}. \label{SiNag}
\end{equation}
Electron-electron (e-e) and electron-phonon (e-ph) scattering affect the noise differently. On one hand, the e-e interaction redistributes the energy of the electron system and $T_e$ increases due to an increasing number of electronic states that contribute to the noise. On the other hand, the power injected by the bias current in the sample can be dissipated to the lattice through e-ph interaction and $T_e$ decreases.\\
In S/N/S junctions, coherent electron-hole pairs penetrate from the superconductor into the normal metal over a distance $L_c=min(L_{\phi},\xi_{\epsilon}=\sqrt{\hbar D/ \epsilon})$ with $\epsilon=max(k_BT,eV)$ and $D$ the diffusion constant of the normal metal. If the sample length is smaller than $L_c$ (coherent case), the phase coherence covers the entire normal region and the Josephson effects determine conductance and noise behavior at low voltage. In this case, the conductance exhibits clear subgap structures (SGS) and the noise is strongly enhanced compared to the normal case due to the coherent transfer of large charge quanta~\cite{SNScoherent,Hoss}.\\
In this paper we consider the incoherent case where $L>>L_c$. This regime has been recently studied theoretically by Bezuglyi et al.~\cite{Bezuglyi} and Nagaev~\cite{Nagaev2001}. They show that the noise is enhanced compared to normal junctions (N/N/N) because of the  confinement of the subgap electrons in the normal part between the superconducting electrodes. If no inelastic processes take place (``collisionless regime'') and in the zero temperature limit the noise increases linearly with the bias voltage:
\begin{equation}
S_I(V)=\frac{2}{3R}(eV+2\Delta), \label{Si}
\end{equation}
where $\Delta$ is the gap of the superconductor.\\
A simplified model to illustrate this behavior is the following: an electron with an energy $eV<<\Delta$ can not escape in the superconducting reservoirs due to the absence of electronic states in the gap. Instead, it enters the superconductor together with a second electron as a Cooper pair and a hole is retroreflected in the normal metal (Andreev reflection). The reflected hole travels the normal region a second time and is retroreflected as an electron at the other S/N interface and so on. In the incoherent case the phase information between two subsequent Andreev reflections is lost. Therefore the quasiparticles experience Incoherent Multiple Andreev Reflections (IMAR). During each travel across the junction, the gain in energy is equal to $eV$, where $V$ is the applied voltage. Therefore the quasiparticles travel the normal part of the junction $N$ times with $N=int[2\Delta /(eV)+1]$ before acquiring enough energy to escape in the superconducting electrode. Within this description, the total noise is the shot noise in a diffusive normal metal at zero temperature $\frac{1}{3}2eI$ times $N$.\\
\begin{figure}[h]
\includegraphics[width=0.49\textwidth,bb=162 142 617 388,clip]{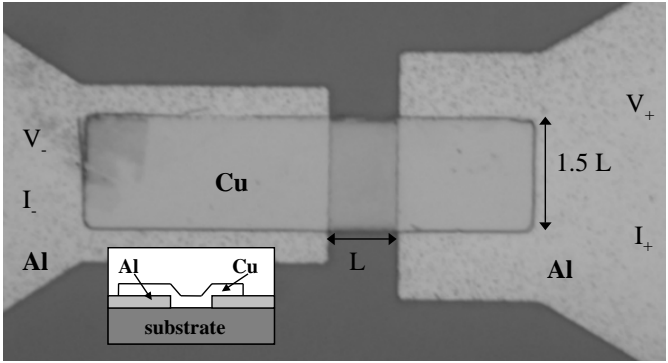}
\caption{Photography of a typical sample (here $L=4\, \mu m$) and schematic cross section. All the dimensions scale with the length $L$.}
\label{fig:photo}
\end{figure}  
At low voltage and finite temperature, the effective length of the junction for the multiply retroreflected particles $L_{eff}=NL\sim L\Delta/V$ exceeds the inelastic length $L_{in}$. In this ``interacting regime'' e-e-collisions interrupt the IMAR cycle before the quasiparticles reach the gap. In the case of strong interaction a Fermi distribution with an effective temperature $T_e$ is restored. $T_e$ decreases with decreasing voltage and reaches equilibrium (lattice temperature) at zero bias. Simultaneously the noise drops from the strongly enhanced level described by Eq. (\ref{Si}) to the Johnson-Nyquist level. Contrary to the N/N/N-case, e-e-interactions reduce the energy window of the involved electronic states.\\
Note that the voltage dependence of the effective length $L_{eff}$ in S/N/S junctions provides a unique way to study inelastic interactions in a normal metal because the same sample can be tuned from the strongly thermalized regime to the collisionless regime simply by changing the applied voltage.\\
\sloppy{On the experimental side only a few results are reported on current noise in incoherent diffusive S/N/S junctions at present. Besides measurements on short (coherent) junctions, Hoss et al.~\cite{Hoss} also addressed the incoherent case. They studied $2\, \mu m$ long Nb/Au/Nb junctions in the interacting regime over the entire voltage range because of the large superconducting gap of Nb. In a paper by Roche et al.~\cite{Roche} the length of the normal part is also about $2\, \mu m$, and despite the use of a high mobility 2DEG as the normal part, the e-ph scattering rate is high enough to drive the junction in the interacting regime. Finally Jehl et al. studied Nb/Al/Nb junctions at relatively high temperatures which behave as two S/N contacts in series~\cite{JehlPRL}. Using a different technique Pierre et al. measured directly the distribution function in long S/N/S junction but only at energies above the gap~\cite{PierrePRL}.\\}
Whereas all these works investigated contacts with a length $L\leq 5\, \mu m$, we deliberately choose to study longer junctions ($4, 10$ and $60\, \mu m$) to determine the respective role of e-e collisions, e-ph collisions and heat transfer through the S/N interfaces on the shot noise in S/N/S junctions. Moreover, to see the crossover from the interacting regime to the collisionless regime, we used aluminium because of its small superconducting gap. Because of the very small phase coherence length in the normal metal we used (see below), we have an almost perfect realization of the regime of IMAR.\\
To measure the current fluctuations we used a SQUID-based experimental setup ~\cite{JehlRSI}. The intrinsic noise is about $10\, \mu \Phi_0 /\sqrt{Hz}$ which is equivalent to $2\, pA /\sqrt{Hz}$ in the input coil of the SQUID. The same experimental setup has been used to perform transport measurements.\\
Samples were fabricated by DC magnetron sputtering and optical lithography. First, a bilayer of Al/Cu ($130\, nm/30\, nm$) is deposited in situ to ensure a good contact between the two metals. This bilayer is etched to define the electrodes. Then the copper bridge (thickness $130\,nm$, purity of the Cu target: $99.9999\%$) is deposited by lift-off, preceded by a short backsputtering to clean the copper surface. Finally the whole sample is etched to remove the copper film from the aluminium electrodes. The resulting thickness of the copper bridge is $90\, nm$. A typical sample and a schematic cross section are shown in Fig.~\ref{fig:photo}.\\
We studied junctions with 3 different lengths $L$: $4$, $10$ and $60\, \mu m$, but with the same width/length ratio: $w/L \approx 3/2$. All samples originate from the same wafer. In spite of very different overlap surfaces between aluminium and copper (because all dimensions scale with the length $L$), all samples have roughly the same resistance of about $0.65\, \Omega$, which indicates a good interface (small barrier resistance). Below the transition of the Al electrodes at $1.5\, K$ and the transition of the Al/Cu bilayer (overlap region) at about $1\, K$ (see right inset of Fig.~\ref{fig:r(v)}), the measured resistance is therefore essentially that of the normal part of the junctions. We then deduce the diffusion constant in copper: $D=30\, cm^2 s^{-1}$. We confirmed this value by measuring the resistance of a meander line consisting of $700$ squares in series, which was cosputtered on the same wafer.\\
In the temperature range from $1\, K$ to $0.3\, K$ the resistance decreases as expected in the classical proximity-effect. This fact and the reduced transition temperature of the bilayer are other indications of clean S/N interfaces. The resistance does not go down to zero since the sample lengths are much greater than the thermal length $L_T=\sqrt{\frac{\hbar D}{k_BT}} \approx 0.9\, \mu m$ at $T=30\, mK$. Therefore no supercurrent is observed. On the contrary, at low temperatures ($T<0.3\, K$), the resistance increases again. This behavior is very similar to the reentrance in S/N junctions first observed by Charlat et al.~\cite{Charlat}. It means the two S/N interfaces of the junctions are connected incoherently. The temperature $T_r$ at which the resistance behavior changes is related to the phase breaking length $L_{\phi}$ by $k_BT_r\approx \hbar D/L_{\phi}^2$ \cite{Courtois} (leading to $L_{\phi}\approx 0.3\mu m$) and is independent of $L$ (see left inset of Fig.~\ref{fig:r(v)}). Such a large phase breaking rate is usual for Cu layers probably because of paramagnetic centers in the Cu oxide at the surface ~\cite{Vranken,Gougam}. The relative amplitude of the reentrance increases as the length $L$ decreases since the relative volume of the sample which is affected by coherent pairs increases.\\
The voltage dependence of the differential resistance $dV/dI$ measured at $T=100\, mK$ (Fig.~\ref{fig:r(v)}) confirms the reentrance behavior. As for the temperature dependence, the amplitude and width of the reentrance peak depend on the width of the sample. However, we clearly see that the voltage needed to destroy the effect ($2\, \mu V$ and $4\, \mu V$ for the $4\, \mu m$ and $10\, \mu m$ samples) is much smaller than $\frac{k_B}{e}T_r$. As suggested in other experiments~\cite{Courtois,Black}, this apparent discrepancy is due to a heating effect. Because we performed conductance and noise measurements simultaneously, we know the electronic temperature at low voltage. Anticipating the detailed description of noise measurements below, we found that the electron gas reaches a temperature of $0.23\, K$ at $2\, \mu V$ for the $4\, \mu m$ sample and $0.2\, K$ at $4\, \mu V$ for the $10\, \mu m$ sample. Consequently the rapid destruction of the reentrance peak is essentially due to the dramatic increase of $T_e$ at low bias voltage.\\
At higher voltage ($V\approx 70\, \mu V$) the bilayer (or at least part of it) is driven in the normal state, and a large peak occurs in the differential resistance (not shown). $dV/dI$ changes by only approximately $10\%$ up to $V=70\, \mu V$ and over this voltage range, we do not observe subharmonic gap structures (SGS) as expected in our experimental situation.\\
SGS appear at voltages $V=2\Delta/ne$ due to the singularity in the DOS of the superconductor at the gap edges and, especially for junctions with high transparent interfaces, to the strongly enhanced probability of Andreev reflection at low energy. The latter is caused by the proximity effect in the normal metal~\cite{Bezuglyi2000}. In our samples the DOS singularities are smeared by the bilayer structure of the electrodes (Al/Cu with good interface) and the proximity corrections are very small because of the very short correlation length $L_c$. Moreover the SGS of high order ($n\geq 4$ because $70\, \mu eV\approx \Delta/2$) are usually very weak and additionally smeared out by inelastic scattering at low voltage.\\      
The results of the noise measurements at $T=100\, mK$ are shown in Fig.~\ref{fig:sr(v)} where we have plotted the current noise density $S_I$ times the resistance $R=V/I$ versus DC voltage. The noise increases more rapidly with the bias voltage than what is expected for two independent S/N junctions in series with a reservoir in between (see dot-dashed line in Fig.~\ref{fig:sr(v)}), and we do not see the thermal crossover towards the Johnson-Nyquist noise level at $eV\approx k_BT \approx 9\, \mu V$. The noise enhancement in these S/N/S junctions is due to the confinement of the quasiparticles between the two superconducting electrodes. With increasing sample length the confinement is relaxed by e-ph interaction and the noise slope at low voltage becomes less important.\\ 
At $ V\approx 50\, \mu V$, the noise of the $4\, \mu m$ sample approaches a straight line compatible with the prediction in the collisionless regime (Eq. (\ref{Si})) with $\Delta=135\, \mu eV$. The gap value is reduced compared to the pure Al film ($\Delta \approx 200\, \mu eV$) because of the bilayer structure of the superconducting electrodes. The reduction factor is the same as for the transition temperature.\\
At higher voltages ($V\geq 70\, \mu V$) the noise shows an irregular behavior which reflects the transition of the bilayer. In this voltage range, all IMAR cycles end by injecting a quasiparticle above the gap in the superconductor. Consequently, many quasiparticles arrive in the superconducting electrodes and weaken the superconductivity. Therefore, the voltage driven transition of the bilayer could be related to the collisionless regime itself.\\
In the following we focus on the interacting regime at low voltage where the electron gas can be described by a Fermi distribution function with an effective temperature $T_e$. Unlike normal junctions where the injected power can be evacuated in the reservoirs, in S/N/S junctions only quasiparticles with energy above the gap can dissipate power. When the sample length is long enough, some heat can be transferred to the phonons through e-ph-scattering. These two mechanisms are covered with our experiments since we studied long to very long junctions.\\
For the longest junction ($60\, \mu m$) we suppose cooling by phonons to be dominant. The electron temperature $T_e$ is then nearly constant over the whole sample length and can be calculated by a heat-diffusion equation~\cite{Henny,Wellstood}. It yields:
\begin{equation}
T_e=(\frac{P}{\Sigma \Omega}+T_{ph}^5)^{1/5} \label{e-ph},
\end{equation}
where $P$ is the power injected in the sample, $\Omega$ its volume and $T_{ph}$ the temperature of the phonon bath (equal to the temperature of the mixing chamber). The parameter $\Sigma$ is the e-ph coupling constant~\cite{Wellstood}. The noise is then given by $4k_BT_e/R$ and we obtain the best fit (see Fig.~\ref{fig:sr(v)} and~\ref{fig:pns}) for $\Sigma = 2.4\, 10^9 Wm^{-3}K^{-5}$, a value which is of the same order of magnitude as reported for Au, Ag, Cu and AuCu~\cite{Henny,PierrePRL,Roukes,Wellstood}.\\
For the shortest junctions ($4\, \mu m$) e-ph-scattering is not very efficient because the volume $\Omega$ is much smaller and the main cooling mechanism is the heat transfer by quasiparticles through the S/N interfaces outside the gap region. Following Bezuglyi et al.~\cite{Bezuglyi} the noise at low voltage is given by the Nyquist formula with a cut off at $T_{e}$ of the order of $\Delta$:
 \begin{equation}
S_I=\frac{4k_BT_e}{R}(1-2exp[-\frac{\Delta}{k_BT_e}]).\label{N_Bez}
\end{equation}
Here the reservoirs are supposed at $T=0\, K$. In fact, we can neglect the finite temperature of the electrodes because the quasiparticle temperature is rapidly much larger than the base temperature. From the measured noise $S_I(V)$ we obtain the effective electron temperature in the normal metal as a function of voltage $T_e(V)$ solving Eq. (\ref{N_Bez}) numerically with $\Delta=135\, \mu eV$.\\
Fig.~\ref{fig:pns} shows the total power $V^2/R$ as a function of the effective temperature of the quasiparticles. For the $4\, \mu m$ sample, we then compare the experimental results to the theoretical prediction obtained by Bezuglyi et al.~\cite{Bezuglyi} who derived the power dissipated through the S/N interfaces as a function of the effective electron temperature at small voltage $eV<<\Delta$ in good agreement with numerical simulations by Nagaev~\cite{Nagaev2001}. The fit in Fig.~\ref{fig:pns} (thick solid line) is given by:
\begin{equation}
P_{NS}\approx \frac{V^2}{R}=\frac{k_BT_e\Delta}{e^2W_{\epsilon}R}\left( 1+\frac{k_BT_e}{\Delta} \right) exp[-\frac{\Delta}{k_BT_e}]. \label{Pns}
\end{equation}
with $W_{\epsilon}=\tau_{ee}(\Delta)/\tau_D$ where $\tau_{ee}(\Delta)$ is the e-e-scattering time at the gap energy and $\tau_D$ the diffusion time. We obtain good agreement between the experimental data and the prediction with $W_{\epsilon}=0.7\pm 0.2$. This value is of the same order as the theoretical estimation of $W_{\epsilon}$ ($W_{\epsilon}=2.3$) using the standard theory of e-e-interaction in a 2D geometry~\cite{Altshuler}. A small but finite interface resistance would renormalize this estimation~\cite{Bezuglyi2000} and could give a better agreement. The fact that $W_{\epsilon} \approx 1$ indicates that the e-e-scattering time at the gap energy is of the order of the diffusion time $\tau_D\approx 5\, ns$ and is two orders of magnitude larger than the phase coherence time found above ($L_{\phi}\approx 0.3\, \mu m$ corresponds to $\tau_{\phi}\approx 0.03\, ns$).\\
Note that the theoretical model applied here is especially dedicated to incoherent S/N/S junctions and therefore more appropriate than the usually used Blonder-Tinkham-Klapwijk (BTK) model~\cite{BTK,Hoss,QuirionJLT} and its derivatives~\cite{Bardas}.\\
For comparison we also plotted in Fig.~\ref{fig:pns} the power $P_{ph}$ dissipated by the phonons in the $4\, \mu m$ sample using the value of $\Sigma$ obtained for the $60\, \mu m$ sample (thin solid line). In the range $200$ to $600\, mK$, $P_{ph}$ is about five times smaller than $P_{NS}$ and the error that we make neglecting this contribution is covered by the uncertainty on $W_{\epsilon}$ given above.\\
Concerning the intermediate sample of length $10\, \mu m$ the power dissipated by phonons and through the interfaces is of the same order of magnitude. Extracting $T_e$ from Eq. (\ref{N_Bez}), we fit the total power by adding the contributions of the two cooling mechanisms, treated separately according to Eq. (\ref{e-ph}) and (\ref{Pns}). We obtain good agreement (see dashed line in Fig.~\ref{fig:pns}) with the following parameters: $\Delta=135\, \mu eV$, the same value of $\Sigma$ as for the $60\, \mu m$ sample and $W_{\epsilon}=0.3 \pm 0.1$ (theoretical estimation $W_{\epsilon}\approx 0.4$).\\        
In conclusion, we investigated IMAR enhanced current noise in long S/N/S junctions of very different lengths ($4, 10$ and $60\, \mu m$). We found that the noise temperature increases very rapidly at low voltage. We deduce the energy dependence of the thermal conductivity of the S/N interfaces which is in good agreement with recent semiclassical theory~\cite{Bezuglyi,Nagaev2001}. The noise behavior of the longest sample can be well fitted taking into account only phonon cooling. The inelastic scattering times we deduced are in agreement with standard description of e-e and e-ph interaction. With the same experimental setup we performed transport measurements. They reveal that in all the samples the two S/N interfaces are connected incoherently, indicating $L_{\phi}<<L_{in}$. The complementarity of transport and noise measurements provided a direct analysis of the voltage dependence of the reentrance in terms of an effective electron temperature.
\begin{acknowledgement}
We would like to acknowledge E.V. Bezuglyi and V.S. Shumeiko for valuable discussions and J.L. Thomassin for technical support.
\end{acknowledgement}

\end{document}